\documentclass[11pt]{article}

\usepackage[margin=1.1in]{geometry}
\usepackage{amsmath,amssymb,amsthm}
\usepackage{algorithm}
\usepackage{algpseudocode}
\usepackage{enumitem}
\usepackage[colorlinks=true,linkcolor=blue,citecolor=blue,urlcolor=blue]{hyperref}

\newtheorem{theorem}{Theorem}[section]
\newtheorem{lemma}[theorem]{Lemma}

\theoremstyle{definition}
\newtheorem{definition}[theorem]{Definition}
\newtheorem{example}[theorem]{Example}

\newcommand{\SetS}{\mathcal{S}}

\newcommand{\TreeEval}{\mathsf{TreeEval}}

\newcommand{\vca}{\mathbf{a}}

\newcommand{\iv}[2]{[#1,#2]}

\newcommand{\T}{\mathcal T}

\newcommand{\val}{\mathsf{val}}

\newcommand{\Mat}[1]{\mathbf{#1}}

\usepackage{mathtools}
\usepackage{cleveref}
\usepackage[normalem]{ulem}

\title{On Pearl Blockchain’s Proof of Useful Work:\\Systematization and Flexibility}

\date{}
\makeatletter
\let\old@fnsymbol\@fnsymbol
\renewcommand{\@fnsymbol}[1]{%
  \ifnum#1=1\relax
    \ensuremath{\dagger}%
  \else
    \old@fnsymbol{#1}%
  \fi
}
\makeatother
\author{Youlong Ding\thanks{The Hebrew University of Jerusalem, Israel. Email: \texttt{youlong.ding@mail.huji.ac.il}.}}

\begin{document}
\maketitle

\begin{abstract}
We revisit the proof-of-useful-work (PoUW) scheme in Pearl blockchain's white paper (Pearl Research Team), giving a systematic sensitivity characterization as an alternative to its Jackpot policy, along with efficient algorithm for verifying the sensitivity condition. 

Beyond its independent interest, our characterization yields a more flexible PoUW scheme that eenables miner autonomy: each miner can independently choose its noise magnitude to tailor the trade-off between computational accuracy and mining efficiency to its own needs.
\end{abstract}

\newpage


\tableofcontents
\newpage

\section{Introduction}

We assume that the reader is familiar with Pearl blockchain's
proof-of-useful-work (PoUW) protocol, as described in its
white paper~\cite{Pearl2026} (also see \cite{komargodski2025proofs}).

In this paper, we focus on the \emph{Jackpot
policy} in~\cite[Section 5]{Pearl2026}, which aims to prevent miners from exploiting
adversarially chosen inputs to perform the prescribed matrix
multiplication substantially faster than a generic computation.

At a high level, the Jackpot policy consists of four checks:
\begin{enumerate}
    \item \emph{Entry liveness.}
    Limits the fraction of input entries whose magnitudes
    overwhelm the injected noise, making their quantized
    values largely predictable.

    \item \emph{Noise floor.}
    Requires every input row to contain a sufficient amount
    of noise in quantized units.

    \item \emph{Tamed products.}
    Limits the fraction of output entries whose accumulation
    magnitudes are unusually large relative to the noise.

    \item \emph{Unpredictable summands.}
    Requires most summands to have sufficient variability
    relative to the accumulator's numerical precision.
\end{enumerate}

While these checks address plausible attack vectors, we
identify two limitations:
\begin{itemize}
    \item  First, the policy is rather
\emph{ad hoc}: it consists of several separate numerical
conditions with carefully chosen thresholds, rather than
a unified principle. This makes the policy difficult to
understand and its completeness difficult to assess.
    \item Second, the policy prescribes a fixed relative noise
magnitude, leaving miners little freedom to tailor the
noise to their individual needs. In particular, different
miners may prefer different trade-offs between computational
accuracy and mining efficiency.
\end{itemize}

\paragraph{Our contributions.}
We propose a simple and systematic \emph{sensitivity condition}
as an alternative way to characterize the effect of injected
noise. Informally, the condition requires that removing any
single entry of either noise matrix, while keeping all other
entries unchanged, must change the resulting matrix product.
We formalize this condition over general finite types,
allowing non-associative arithmetic such as floating-point
accumulation.

We then give an efficient algorithm for verifying the
sensitivity condition. For locally monotone operations,
our algorithm runs in
$O(n^3\log|\SetS|)$ time, nearly matching the cost of
a single matrix multiplication. The key idea is to propagate
intervals backward through the evaluation tree, allowing
the effect of every noise entry to be tested without
recomputing the matrix product.

Beyond its independent theoretical interest, our
characterization also enables a more \emph{flexible}
PoUW design through \emph{miner autonomy}. Rather than
imposing a uniform noise magnitude, miners can independently
choose their noise levels according to their own priorities.
Smaller noise improves computational accuracy but may reduce
the likelihood of satisfying the sensitivity condition,
whereas larger noise may improve this likelihood at the
cost of greater computational error.

\section{Preliminaries}

\begin{definition}[Chain]
A  \emph{chain} is a finite totally ordered set
\[
 \SetS = \{s_1, \dots, s_q\},\qquad s_1 < \cdots < s_q, \qquad q\ge1.
\]
For $L, U \in \SetS$, $L \leq U$, define its interval
$$
\iv{L}{U} \coloneqq \{s\in\SetS:L\leq s\leq U\}.
$$
\end{definition}

\begin{definition}[Type]\label{def:finite-type}
A \emph{type} is a triple $(\SetS,\oplus,\otimes)$,
where $\SetS$ is a finite chain and
$\oplus,\otimes:\SetS\times\SetS\to\SetS$
are binary operations satisfying the following properties:
\begin{enumerate}
    \item \emph{Commutativity:}
    For all $a,b\in\SetS$,
    \[
    a\oplus b=b\oplus a,
    \qquad
    a\otimes b=b\otimes a.
    \]

    \item \emph{Zero element:}
    There exists a zero element $0\in\SetS$ such that,
    for every $a\in\SetS$,
    \[
    a\oplus 0=a,
    \qquad
    a\otimes 0=0.
    \]
\end{enumerate}
\end{definition}

\begin{example}
Floating-point arithmetic, e.g., $(\mathsf{FP8}, +, \times)$, $(\mathsf{FP16}, +, \times)$ is a type.
\end{example}

\begin{definition}[Tree evaluation]
Let $\oplus: \SetS \times \SetS \to \SetS$
be a binary operation on a chain $\SetS$.
Let $\T$ be a binary tree with $n$ leaves.
Given $\vca = (a_1,\dots,a_n) \in \SetS^n$,
we assign the value $a_i$ to the $i$-th leaf.
We recursively define the value $\val(v)$ of each node $v$:
\[
\val(v) \coloneq
\begin{cases}
a_i, & \text{if $v$ is the $i$-th leaf},\\
\val(v_L) \oplus \val(v_R),
& \text{if $v$ has left child $v_L$ and right child $v_R$}.
\end{cases}
\]
The \emph{tree evaluation} is the value at the root:
\[
\TreeEval_{\T}(\vca,\oplus)
\coloneq \val(\operatorname{root}(\T)).
\]
\end{definition}

\begin{definition}[$n$-ary extension of a binary operation]
Let $(\SetS,\oplus,\otimes)$ be a type, and let
$\T$ be a binary tree with $n$ leaves.
We define the $n$-ary extension of $\oplus$ induced by $\T$ as
\[
\bigoplus\nolimits_{\T}\vca
\coloneq \TreeEval_{\T}(\vca,\oplus),
\qquad \vca\in\SetS^n.
\]
\end{definition}

\begin{definition}[Local monotonicity]\label{def:local-mono}
Let $\SetS$ be a chain, $f:\SetS^n\to\SetS$.
We say that $f$ is \emph{locally monotone} if, for every
$i\in[n]$ and every fixed choice of $x_j\in\SetS$ for $j\neq i$,
the unary map
\[
f_i:\SetS\to\SetS,\qquad
t\mapsto f(x_1,\dots,x_{i-1},t,x_{i+1},\dots,x_n)
\]
is either nondecreasing or nonincreasing.
\end{definition}

\begin{definition}[Matrix multiplication]\label{def:matmul}
Let $(\SetS,\oplus,\otimes)$ be a type,
and let $\T$ be a binary tree with $n$ leaves.
For $\Mat A,\Mat B\in\SetS^{n\times n}$,
we define their \emph{matrix product}
$\Mat A\cdot_{\T}\Mat B\in\SetS^{n\times n}$ by
\[
(\Mat A\cdot_{\T}\Mat B)_{i,j}
\coloneq \bigoplus\nolimits_{\T}
\bigl(\Mat A_{i,1}\otimes\Mat B_{1,j},
\dots,
\Mat A_{i,n}\otimes\Mat B_{n,j}\bigr)
\]
for all $i,j\in[n]$.
When $\T$ is understood, we simply write
$\Mat A\cdot\Mat B$.
\end{definition}


\section{Systematic Characterization: Sensitivity Condition}

\begin{definition}[Sensitivity condition]\label{def:sensitivity}
Let $(\SetS,\oplus,\otimes)$ be a type.
For $\Mat A,\Mat B,\Mat E_1,\Mat E_2\in\SetS^{n\times n}$,
let
\[
\Mat C\coloneqq
(\Mat A\oplus \Mat E_1)\cdot (\Mat B \oplus \Mat E_2),
\]
where $\oplus$ is naturally extended by applying it entrywise.
We say that the \emph{sensitivity condition} holds for
$(\Mat A,\Mat B,\Mat E_1,\Mat E_2)$ with respect to $\T$
if, for every $(i,j)\in[n]^2$,
\[
(\Mat A+\Mat E_1^{\backslash(i,j)})
\cdot(\Mat B+\Mat E_2)
\neq\Mat C,
\qquad
(\Mat A+\Mat E_1) \cdot
(\Mat B+\Mat E_2^{\backslash(i,j)})
\neq\Mat C,
\]
where  $\Mat E^{\backslash(i,j)}$ denote the matrix obtained
from $\Mat E$ by replacing its $(i,j)$-th entry with $0_{\SetS}$.
\end{definition}

\section{Efficient Verification of Sensitivity Condition}
\label{sec:efficient-verification}

There are $2n^2$ possible noise-entry removals. Recomputing a
full matrix product for each takes $O(n^5)$ time; even recomputing
only the affected row or column takes $O(n^4)$ time. We show that
local monotonicity of $\oplus$ allows all tests to be performed in
$O(n^3(1+\log|\SetS|))$ time.

\begin{theorem}[Efficient verification]\label{localalg:main}
Let $(\SetS,\oplus,\otimes)$ be a type, and let $\T$ be a binary
tree with $n$ leaves. If $\oplus$ is locally monotone, then the
sensitivity condition for any
$\Mat A,\Mat B,\Mat E_1,\Mat E_2\in\SetS^{n\times n}$
with respect to $\T$ can be verified in
$O(n^3(1+\log|\SetS|))$ time, assuming constant-time operations
on $\SetS$ and indexed access to its ordered elements.
\end{theorem}

\subsection{Inverting a monotone map}

Our basic tool is an interval inversion procedure that does not need
to know the direction of monotonicity.

\begin{lemma}[Interval inversion]\label{localalg:inverse}
Let $f:\SetS\to\SetS$ be either nondecreasing or nonincreasing,
and let $I\subseteq\SetS$ be an interval. Suppose that
$x_0\in\SetS$ satisfies $f(x_0)\in I$.
Then $f^{-1}(I)$ is an interval containing $x_0$, and its endpoints
can be found using $O(1+\log|\SetS|)$ evaluations of $f$.
\end{lemma}

\begin{proof}
For $x\le y\le z$, monotonicity implies that $f(y)$ lies
between $f(x)$ and $f(z)$. Thus $f(x),f(z)\in I$ implies
$f(y)\in I$, so $f^{-1}(I)$ is an interval.
Since $x_0\in f^{-1}(I)$, the predicate $f(x)\in I$ is false
and then true to the left of $x_0$, and true and then false to its
right. Two binary searches on the ordered elements of $\SetS$
therefore find the endpoints, regardless of the monotonicity
direction.
\end{proof}

\subsection{Backward propagation on the evaluation tree}

Write
\[
\Mat X\coloneq\Mat A\oplus\Mat E_1,\qquad
\Mat Y\coloneq\Mat B\oplus\Mat E_2,\qquad
\Mat C\coloneq\Mat X\cdot_{\T}\Mat Y,
\]
where matrix addition is entrywise. Fix an output entry $(i,j)$.
Its evaluation tree $\T$ has leaf values
\[
p_k\coloneq\Mat X_{i,k}\otimes\Mat Y_{k,j},
\qquad k\in[n],
\]
and root value $\Mat C_{i,j}$. Evaluate the tree once and store
$\val(v)$ for every node $v$.

For each node $v$, let $I_v\subseteq\SetS$ be the set of values
that can replace $\val(v)$ without changing the root value,
while all other parts of the computation remain fixed.
In particular,
\[
I_{\operatorname{root}(\T)}=\{\Mat C_{i,j}\}.
\]
If $v$ has left child $u$ and right child $w$, then
\begin{equation}\label{localalg:propagate}
\begin{aligned}
I_u&=\{t\in\SetS:t\oplus\val(w)\in I_v\},\\
I_w&=\{t\in\SetS:\val(u)\oplus t\in I_v\}.
\end{aligned}
\end{equation}
Because $\oplus$ is locally monotone, each set is the inverse
image of an interval under a monotone unary map. Moreover,
$\val(u)\in I_u$ and $\val(w)\in I_w$.
By Lemma~\ref{localalg:inverse}, both child intervals can
therefore be computed in $O(1+\log|\SetS|)$ time.
Propagating these intervals from the root to the leaves gives,
for each $k\in[n]$, the interval $I_k$ of leaf values that
preserve $\Mat C_{i,j}$.

Now consider removing one noise entry. Replacing
$(\Mat E_1)_{i,k}$ by $0$ changes only the $k$-th leaf of this
output computation, whose new value is
\[
p^{(1)}_{i,j,k}\coloneq\Mat A_{i,k}\otimes\Mat Y_{k,j}.
\]
Similarly, replacing $(\Mat E_2)_{k,j}$ by $0$ gives
\[
p^{(2)}_{i,j,k}\coloneq\Mat X_{i,k}\otimes\Mat B_{k,j}.
\]
Consequently, the output entry $(i,j)$ changes under the first
or second removal exactly when, respectively,
\begin{equation}\label{localalg:leaf-tests}
p^{(1)}_{i,j,k}\notin I_k
\qquad\text{or}\qquad
p^{(2)}_{i,j,k}\notin I_k.
\end{equation}
No monotonicity assumption on $\otimes$ is needed: each candidate
leaf value is computed directly.

\subsection{Verification algorithm and analysis}

Initialize two $n\times n$ arrays of counters, $d^{(1)}$ and
$d^{(2)}$, to zero. For each $(i,j)\in[n]^2$, evaluate the
corresponding tree and propagate the intervals as in
\eqref{localalg:propagate}. For every leaf $k$, increment
$d^{(1)}_{i,k}$ if $p^{(1)}_{i,j,k}\notin I_k$, and increment
$d^{(2)}_{k,j}$ if $p^{(2)}_{i,j,k}\notin I_k$.
The algorithm accepts if and only if every counter is positive.

\begin{proof}[Proof of Theorem~\ref{localalg:main}]
The intervals computed by backward propagation have exactly the
stated meaning. This holds at the root, and
\eqref{localalg:propagate} preserves it when passing from a
parent to either child while fixing the other child's value.
Thus \eqref{localalg:leaf-tests} correctly determines whether removing
a noise entry changes the output $(i,j)$.

Removing $(\Mat E_1)_{i,k}$ can affect only row $i$ of the
matrix product. Hence $d^{(1)}_{i,k}>0$ exactly when this removal
changes the output matrix. Likewise, removing
$(\Mat E_2)_{k,j}$ can affect only column $j$, and it changes
the output matrix exactly when $d^{(2)}_{k,j}>0$.
Therefore all counters are positive precisely when the
sensitivity condition holds.

Each output entry requires a forward evaluation of $O(n)$ nodes,
a backward traversal with $O(1+\log|\SetS|)$ work per node,
and $O(n)$ leaf tests. Over all $n^2$ output entries, this takes
$O(n^3(1+\log|\SetS|))$ time. The matrices and counters use
$O(n^2)$ space, and the values and intervals for one tree use
$O(n)$ additional space.
\end{proof}

\section{PoUW Scheme Made Flexible}
Beyond its independent interest, our systematic characterization enables a desirable feature of the resulting PoUW scheme: miner autonomy. Namely, rather than imposing a uniform noise magnitude rule as in \cite{Pearl2026}, the resulting PoUW scheme allows each miner to choose its own noise level, tailoring the trade-off between computational accuracy and mining efficiency to its individual needs:
\begin{enumerate}
    \item \emph{Smaller noise.}
    A miner prioritizing accuracy may choose smaller noise,
    reducing the deviation from the original matrix product,
    but potentially lowering the probability that the
    sensitivity condition holds.

    \item \emph{Larger noise.}
    A miner prioritizing mining efficiency may choose larger
    noise, potentially increasing the probability of satisfying
    the sensitivity condition at the cost of greater
    computational error.
\end{enumerate}


\section*{Acknowledgements}
Ding was supported in part by a grant from the Israel Science Foundation (ISF Grant No. 1774/20), and by the European Union (ERC, SCALE,101162665). 
Views and opinions expressed are
however those of the author(s) only and do not necessarily reflect those of the
European Union or the European Research Council. Neither the European Union
nor the granting authority can be held responsible for them.

\bibliographystyle{alpha}
\bibliography{ref}

\end{document}